# Interdependency between the Stock Market and Financial News

**EunJeong Hwang[1] and Yong-Hyuk Kim[2]**

[1]College of Information and Computer Sciences, University of Massachusetts, Amherst
[2]School of Software, Kwangwoon University, Seoul, Republic of Korea
ehwang@umass.edu, yhdfly@kw.ac.kr

## Abstract

Stock prices are driven by various factors. In particular, many individual investors who have relatively little financial knowledge rely heavily on the information from news stories when making investment decisions in the stock market. However, these stories may not reflect future stock prices because of the subjectivity in the news; stock prices may instead affect the news contents. This study aims to discover whether it is news or stock prices that have a greater impact on the other. To achieve this, we analyze the relationship between news sentiment and stock prices based on time series analysis using five different classification models. Our experimental results show that stock prices have a bigger impact on the news contents than news does on stock prices.

## Introduction

News and stocks tend to be very sensitive to social issues. Recently, predicting stock price movements using news articles has attracted the interest of many researchers (Kalyani et al., 2016). However, we may only hear good news when stocks have already peaked. Moreover, stocks often rebound shortly after a series of bad news. In this paper, we analyze the impact of stock prices on news as well as the impact of news on stock prices, and we clarify correlation between them. We opted to consider North Korean stocks and Tesla, Inc. as our two stock market subjects. We choose them mainly because they have recently shown great volatility. Highly volatile stocks can facilitate our understanding of the relationship between stocks and news over the short term.

The key events in North Korea that were behind its stock market fluctuations were the Demilitarized Zone meeting between US President Donald Trump and North Korean Leader Kim Jong Un in July 2019 and a ballistic missile launch in May 2019. In the case of Tesla, the biggest sales drop in its history occurred due to the failure of supply management for China and Europe in April 2019. In contrast, in June 2019, Tesla's stock rebounded sharply as it set a new record for deliveries in North America.

In this paper, we use supervised machine learning models to study the relationship between news and stocks. We analyze the correlations between them through classification and regression by applying each model to the time series data of news sentiments and stock prices. To extract important terms and sentiments from news stories, we utilize a variety of natural language processing and text mining techniques. Using the correlation between news and stocks obtained from the above techniques, we identify which responds more quickly to social issues and how they affect each other.

## Research Methodology

### News and Stock Data Collection

We collected news articles about North Korea and Tesla between April 3, 2019 and July 8, 2019. We developed a Web scraper using NewsAPI[1] to obtain the information of each article, which includes its URL, publication date, and headline. In the case of North Korea, we searched for keywords such as "Trump," "missile," and "nuclear" together to obtain news that affected its economic situation. We also limited news sources to 36 United States organizations that publish many financial new stories, such as CNBC, Reuters, and the *Wall Street Journal*. The whole text of each article was extracted from its specific URL using the BeautifulSoup library in Python. Then, we added one day to the date of crawled articles to match the time of the North Korean stock market and the US news sources. Therefore, the North Korean stock prices correspond with articles from the previous day.

The daily stock prices of North Korea and Tesla were extracted by using the Naver and Yahoo Finance application programming interfaces (APIs), respectively. To calculate a share price representing North Korea, we selected five North Korea-related stocks in different industries and averaged their prices. The data set of stock prices contains the opening, high, low, and closing prices as well as volume. For data consistency, the closing prices were used as the general stock prices.

---

[1] https://newsapi.org/docs. NewsAPI is free and a simple HTTP REST API for searching and retrieving live articles from all over the Web. It indexes articles from over 30,000 worldwide sources.

### Preprocessing

To reduce noise and minimize data discrepancy, we preprocessed the text data. Then, we defined the keywords related to North Korea as "North Korea," "N.Korea," and "NKorea" and defined Tesla's keywords as "Tesla" and "Tsla." Figure 1 presents the steps of our text preprocessing.

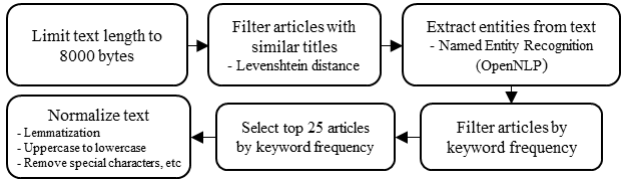

Figure 1: Text pre-processing steps

First, we limited the text length of news articles to remove articles that include excessive information that is not related to keywords. To prevent redundancy, we applied the Levenshtein distance to the title of the articles and removed the articles if titles were similar to each other by more than 80%. Then, the keywords' frequencies in the articles were counted by extracting entities using the named entity recognition function of OpenNLP. For articles on North Korea, we filtered out an article if the keyword was mentioned less than once, and for those on Tesla, the articles that have fewer than three keywords were removed. Based on keyword frequency, the top 25 articles were selected. For the chosen articles, we cleaned the data by removing white spaces and special characters and by converting all uppercase letters to lowercase ones. We also performed lemmatization, replacing all words with their base form using the WordNet lemmatizer in Python.

### Candidate Extraction and Synonym Detection

To identify terms representing the contents of the article, we extracted candidates composed of adjectives and nouns using the part-of-speech tagger in Python's NLTK library. Then, we considered two similarity measures, WordNet and cosine similarity, to identify synonyms of the extracted candidates. To choose a better similarity measure, we evaluated the correlation between the candidates and stock prices for North Korea and Tesla using both methods. The average correlations of WordNet similarity and cosine similarity were –0.068 and 0.129, respectively. Because it had a higher correlation, the second method was used to process synonyms. We used *word2vec* in Python and Google's pre-trained *word2vec* model for word embedding. Using these vector values, we calculated a candidate's vector value using the average of each word vector value. In addition, we removed non-financial candidates through our own stop-word list.

### Important Term Identification and Sentiment Analysis

After extracting candidates, TF-IDF (term frequency-inverse document frequency) was used to identify the important candidates. For sentiment analysis, we considered two measures. The first was to use the sentiment words list created by University of Illinois at Chicago (UIC) group (Hu and Liu, 2004). The other was to use the UIC team's list combined with the Stanford natural language processing tool. The average correlation of the first method was calculated and found to be 0.129, whereas that of the second method was 0.018. Hence, we used the first method to calculate the sentiment of the candidates. We calculated sentiment scores by counting positive and negative words in the candidate as "#(positive words) – #(negative words)" (Taj et al., 2019). If the sentiment score is greater than 0, the text is tagged as "positive" and if it was less than 0, the text was tagged as "negative." Next, we calculated the rate of change of daily sentiment score and stock prices as "100 × (*value*(base date) – *value*(1 day before)) / *value*(1 day before)." To adjust values from different scales to common scale, we normalized all data to range between –1 and 1.

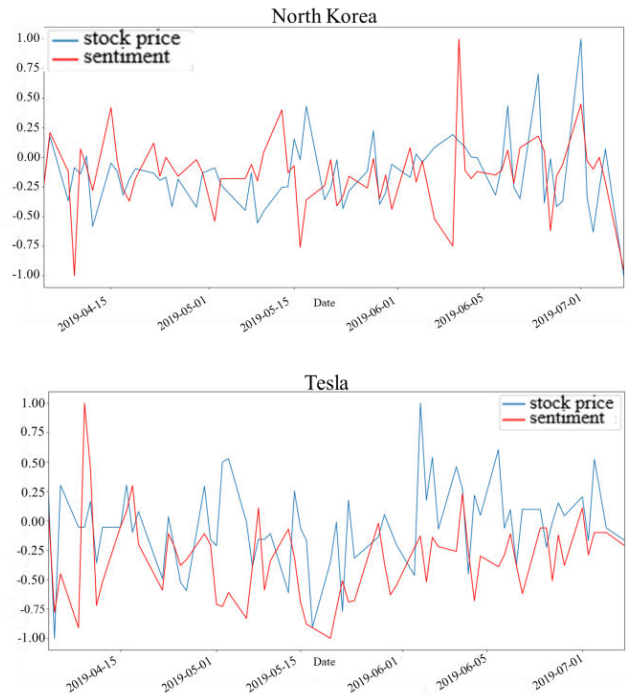

Figure 2: Comparison of stock price and sentiment index variation for (top) North Korea and (bottom) Tesla

Table 1: Comparison of the correlations between stock price variation and news sentiment variation using five classifiers

| | X data range | X → Y | Type | GPs | LR | MLP | SMOreg | RF | Average |
|---|---|---|---|---|---|---|---|---|---|
| **Regression** | Past 5 days + base date + future 1 day | articleToStock | NKorea | 0.178 | 0.280 | 0.224 | 0.086 | 0.163 | 0.186 |
| | | | Tesla | 0.187 | 0.169 | 0.376 | –0.129 | –0.198 | 0.081 |
| | | stockToArticle | NKorea | –0.151 | –0.261 | 0.324 | –0.190 | –0.079 | –0.072 |
| | | | Tesla | 0.476 | 0.113 | 0.228 | 0.327 | 0.459 | 0.321 |
| | Base date | articleToStock | NKorea | 0.308 | 0.000 | 0.406 | –0.061 | 0.114 | 0.153 |
| | | | Tesla | 0.273 | 0.111 | 0.186 | 0.189 | 0.312 | 0.215 |
| | | stockToArticle | NKorea | 0.218 | 0.000 | -0.434 | 0.270 | –0.126 | –0.014 |
| | | | Tesla | 0.508 | 0.508 | 0.406 | 0.208 | 0.575 | 0.441 |
| **Classification** | Past 5 days + base date + future 1 day | articleToStock | NKorea | –0.109 | –0.045 | -0.054 | –0.386 | 0.208 | –0.077 |
| | | | Tesla | –0.213 | –0.430 | -0.300 | –0.247 | –0.199 | –0.278 |
| | | stockToArticle | NKorea | 0.354 | 0.426 | 0.250 | 0.335 | 0.910 | 0.455 |
| | | | Tesla | 0.714 | 0.798 | 0.905 | 0.748 | 0.968 | 0.826 |
| | Base date | articleToStock | NKorea | 0.202 | 0.000 | 0.082 | –0.344 | –0.180 | –0.048 |
| | | | Tesla | 0.144 | 0.067 | 0.236 | 0.106 | –0.007 | 0.109 |
| | | stockToArticle | NKorea | 0.712 | 0.741 | 0.907 | 0.598 | 0.986 | 0.789 |
| | | | Tesla | 0.628 | 0.762 | 0.997 | 0.785 | 1.000 | 0.834 |

## Dataset Generation and Machine Learning

With time series data sets, we divided the analysis into regression and classification to analyze the relationship between stock prices and sentiment. In the regression analysis, we conducted two experiments: in the first experiment, stock price change rate, and trading volumes were used as independent variables, labeled $X$, and the sentiment index change rate was the dependent variable, labeled $Y$; in the second experiment, we used the sentiment index change rate and the number of articles as independent variables and the stock price change rate was the dependent variable. Two experiments were also conducted in the classification problem. The independent variables were set as the above in both experiments. The dependent variable is –1 or 1 when the change rate is less than 0 or greater than 0, respectively. We also scaled the values of trading volumes and the number of articles so that they had the same range [0,1].

For machine learning models, we used Gaussian processes (GPs), linear regression (LR), multilayer perceptron (MLP), support vector regression (SMOreg), and random forest (RF) algorithms suitable for regression and classification problems. All five algorithms were implemented and evaluated using the Weka[2] tool. To determine how many days of independent variables were appropriate for predicting the dependent variables, we tested each model with independent variables of [base date – 5 days ≤ base date ≤ base date + 1 day], [base date – 6 days ≤ base date ≤ base date + 1 day], and [base date – 7 days ≤ base date ≤ base date+ 1 day]. The average correlations for the three ranges on combined North Korea and Tesla data for the regression problem were 0.129, –0.016, and –0.093, respectively. Because the first range showed the highest correlation, the rest of the experiments, including the classification, were conducted with the same range of data. The performance of each model was evaluated by training the first 66% of the instances and testing the remaining instances.

## Results and Analysis

In the regression analysis, the model was evaluated without the number of articles and trading volume to check their effect on the fluctuations of stock prices and news sentiments. If the number of articles and trading volume were included, the average correlation of the five models was 0.129, whereas if they were not included, it dropped to 0.014. In other words, the number of articles is related to the stock price fluctuation, and trading volume is related to news sentiment fluctuation, when we predict the stock prices and news sentiments. In contrast, in the classification analysis for North Korea and Tesla under the same conditions, the average correlation was 0.232 when we included them and 0.295 when we excluded them. This indicates that the number of articles and trading volume have little impact on the classification of stock prices and article sentiment.

Table 1 presents the results of experiments comparing the correlation between stock prices variation and news sentiments variation. The term "articleToStock" means that stock price fluctuation is predicted from the number of articles and their sentiment fluctuation, whereas "stockToArticle" means that sentiment fluctuation is predicted from the stock price fluctuation and trading volume.

---
[2] https://www.cs.waikato.ac.nz/ml/weka

In the regression analysis for North Korea, the articleToStock experiment, which includes past and future dates, indicates that the content of news articles is weakly correlated with stock prices. In the stockToArticle experiment, the average correlation of all models except for MLP was –0.17. This means that North Korean stock prices are also rarely related. In contrast, in the North Korea's articleToStock experiment, when the independent variables contain only the base date values, the GPs and MLP models yield high correlation. The stockToArticle experiment also yields high correlation for the GPs and SMOreg models under the same conditions. Although we did not include other experimental results due to the space limits of this paper, when North Korea's articleToStock and stockToArticle experiments were conducted with only the past five days of $X$ values, and the average correlations were 0.035 and 0.024, respectively. Moreover, for the articleToStock experiment with the base date and one future day as the $X$ value range, the average correlation of all four models except RF was –0.277. This means that when the data of the past and the future are included, the prediction rates of stock prices and sentiments drop. This demonstrates that the stocks and articles reflect North Korea's drastic changes on a daily basis. In particular, as we mentioned in the second section (Research Methodology), the North Korean articles represent the future stock prices. Thus, the news reflects the situation of North Korea more slowly than do stocks. In the Tesla's regression analysis, the average correlation of articleToStock, including past and future dates, was 0.081. This shows that articles are little relevant to stock prices. However, on the base date, GPs and RF models of its articleToStock experiment yield high correlation. Tesla-related articles published on the base date have a relatively higher relevance to the stocks than those published on other days. In contrast, in the Tesla's stockToArticle experiment, including the past and future dates, the GPs, SMOreg, and RF models showed high correlations of 0.476, 0.327, and 0.459, respectively. This means that stock prices have a high impact on news sentiments. In particular, in the case of stockToArticle experiment with the base date, the average correlation is 0.441, indicating that stock prices on the base date also have a great impact on the articles.

In the classification problem for North Korea, the average correlation of the articleToStock experiment including the past and future periods was –0.077 and the average correlation using its base date was –0.048. This means that we cannot predict whether the stock price will rise or fall using news sentiments. However, in the stockToArticle experiment including the past and future periods, the average correlation of the five models was 0.455. Its correlation for base date was especially high as well. This means that stock prices might substantially influence news sentiments. In the articleToStock experiment of Tesla's classification problem, the average correlation including past and future periods was –0.278 and the correlation including only the base date was 0.109. This indicates that articles do not have a big impact on stocks. In contrast, in the stockToArticle experiment, when we included past and future dates, the average correlation was 0.826 and the correlation of the base date was also very high. Therefore, Tesla's stocks might have a big impact on article sentiments.

## Concluding Remarks

In this study, we explored whether stock prices or news article sentiments are more sensitive to social issues as well as whether stock prices or articles have more influence on the other. It was observed that stock prices have a greater impact on the sentiment of the articles. In other words, stock prices responded to social issues before the articles do. In addition, it might be not easy to predict stock prices using news stories. Stock price prediction is even more difficult if past and future articles are included. This is likely to be because North Korea's and Tesla's situations change rapidly on a daily basis. In this study, only three months of articles were collected due to the limitations of the API. To extract candidates, we extracted all phrases consisting of adjectives and nouns. This caused the candidates to contain too many general terms. In future work, we can use text mining tools to extract more sophisticated candidates. We plan to add more articles as well. Based on these approaches, we would like to extend this study by analyzing the impacts of long-term past articles on stock prices or by analyzing the relationship between social network services and stocks.

## References

Minqing Hu and Bing Liu. 2004. Mining and summarizing customer reviews, In Proceedings of the 10th ACM SIGKDD International Conference on Knowledge Discovery and Data Mining. Pages 168-177.

Joshi Kalyani, H. N. Bharathi, Rao Jyothi. 2016. Stock trend prediction using news sentiment analysis. arXiv:1607.01958 [cs.CL]. K. J. Somaiya College of Engineering, Mumbai

Soonh Taj, Baby Bakhtawer Shaikh, Areej Fatemah Meghji. 2019. Analysis of news articles: a lexicon based approach. In Proceedings of the 2nd International Conference on Computing Mathematics & Engineering Technologies. Pages 1-5.